\documentclass[11pt]{iopart}

\usepackage{graphicx}
\usepackage{color}

\newcommand {\bfp} {{\bf p}}
\newcommand {\bfq} {{\bf q}}

\renewcommand {\d} {{\rm d}}

\newcommand {\E} {\varepsilon}

\newcommand {\om} {\omega}
\newcommand {\Om} {\Omega}

\begin{document}

\paper{Interplay of the volume and surface plasmons in the electron energy loss spectra of C$_{60}$}

\author{A V Verkhovtsev$^{1,2}$, A V Korol$^1$, A V Solov'yov$^1$\footnote{On leave from A.F. Ioffe Physical-Technical Institute, St. Petersburg, Russia}, P Bolognesi$^3$, A Ruocco$^4$ and L Avaldi$^3$}

\address{$^1$ Frankfurt Institute for Advanced Studies, Ruth-Moufang-Str. 1, 60438 Frankfurt am Main, Germany}
\address{$^2$ St. Petersburg State Polytechnic University, Politekhnicheskaya ul. 29, 195251 St. Petersburg, Russia}
\address{$^3$ CNR-Istituto di Metodologie Inorganiche e dei Plasmi, Area della Ricerca di Roma 1, CP 10, 00015 Monterotondo Scalo, Italy}
\address{$^4$ Dipartimento di Fisica and Unit\`{a} CNISM, Universit\`{a} di Roma Tre, Via della Vasca Navale 84, 00146 Roma, Italy}

\ead{verkhovtsev@fias.uni-frankfurt.de}


\begin{abstract}
The results of a joint experimental and theoretical investigation of the C$_{60}$ 
collective excitations in the process of inelastic scattering of electrons are presented.
The shape of the electron energy loss spectrum is observed to vary when the scattering angle increases.
This variation arising due to the electron diffraction of the fullerene shell is described 
by a new theoretical model which treats the fullerene as a spherical shell of a finite width
and accounts for the two modes of the surface plasmon and for the volume plasmon as well.
It is shown that at small angles the inelastic scattering cross section is determined mostly 
by the symmetric mode of the surface plasmon, while at larger angles the contributions of the 
antisymmetric surface plasmon and the volume plasmon become prominent.
\end{abstract}



\nosections

The interaction of a charged particle or an electromagnetic field with a many-particle system 
may lead either to the excitation of a single particle state of the system or to the excitation 
of collective states involving many particles.
The latter case is described by the formation of the so-called giant resonances which are 
characterized by the collective motion of charged particles against that of the particles of 
opposite charge. 
Being a general physical phenomenon, this effect has been considered 
in nuclei \cite{Eisenberg_Greiner}, many-electron atoms \cite{Connerade_GR}, 
atomic clusters \cite{deHeer_1993_RevModPhys.65.611,Kreibig_Vollmer} and 
condensed media \cite{Bohm_Pines_1952_PhysRev.85.338,Kaplan_1987_AdvChemPhys.68.225}.

Like condensed media, metal clusters and fullerenes have delocalized electrons which oscillate 
against the positively charged ions forming collective plasmon excitations.
A significant fundamental interest has been aroused in studying the plasmon formation in these systems
\cite{Kreibig_Vollmer,Solovyov_review_2005_IntJModPhys.19.4143,Phaneuf_2010_Plasmon_in_Fullerenes}.
Investigation of potential applications \cite{Singhal_2008_ApplPhysLett.93.103114,Porter_Gass_2008} 
of plasmons formed a new field of physics, named nanoplasmonics.

It is known \cite{Gerchikov_2000_PhysRevA.62.043201} that collective electron excitations in 
metal clusters can be of two different types, namely the surface and the volume plasmons. 
The dipole surface plasmons are responsible for the formation of giant resonances in photoabsorption
spectra of metal clusters, while the volume plasmon modes, which have higher resonance frequencies, 
provide an essential contribution to the formation of the electron impact ionization cross section.

Existence of a giant resonance in the excitation spectra of fullerenes at about 20 eV was predicted 
theoretically \cite{Bertsch_1991_PhysRevLett.67.2690} and then observed experimentally in the case 
of C$_{60}$ in photoionization \cite{Hertel_1992_PhysRevLett.68.784} and inelastic scattering of 
electrons \cite{Keller_Coplan_1992_ChemPhysLett.193.89}. 
Recent experiments on photoionization of neutral \cite{Reinkoester_2004_JPhysB.37.2135} and charged \cite{Scully_2005_PhysRevLett.94.065503} C$_{60}$ molecules revealed the existence of the second 
collective resonance at about 40 eV which firstly was assigned to the volume plasmon \cite{Scully_2005_PhysRevLett.94.065503}. 
Then its assignment to the second surface plasmon has been proposed 
\cite{Korol_AS_2007_PhysRevLett_Comment} and discussed \cite{Scully_2007_PhysRevLett_Reply}.

Theoretical investigations of the scattering of fast electrons on fullerenes \cite{Gerchikov_1997_JPhysB.30.4133,Gerchikov_1998_JPhysB.31.3065} within the single-plasmon 
model predicted the existence of diffraction phenomena, which were then experimentally observed 
in C$_{60}$ in \cite{Mikoushkin_1998_PhysRevLett.81.2707}.

In this paper, we reveal for the first time the contribution and the interplay of the three 
plasmons to the inelastic scattering cross section of electrons on the C$_{60}$ fullerene.
As opposed to the photoionization, the electron impact ionization causes the formation not only 
of two surface plasmons but the volume plasmon as well. 
We show, both experimentally and theoretically, that the volume plasmon manifests itself as the 
scattering angle increases.


We use a simple but physically reasonable model
\cite{PuskaNiemenen_1993_PhysRevA.47.1181,Oestling_1993_EurophysLett.21.539,
Ruedel_2002_PhysRevLett.89.125503,Lambin_Lukas_1992_PhysRevB.46.1794,Lo_2007_JPhysB.40.3973} 
which treats the fullerene as a spherical shell of a finite width, $\Delta R = R_2 - R_1$ 
(where $R_{1,2}$ are the inner and the outer radii of the molecule, respectively).
Interaction with an incident electron leads to the variation of the volume electron density, 
occurring inside the shell, and of the surface electron densities at the inner and the outer 
surfaces of the shell.
These variations lead to the formation of the volume plasmon \cite{Gerchikov_2000_PhysRevA.62.043201} 
and two coupled modes of the surface plasmon, a symmetric and an antisymmetric one \cite{Oestling_1993_EurophysLett.21.539,Lambin_Lukas_1992_PhysRevB.46.1794,Lo_2007_JPhysB.40.3973}.
Within the plasmon resonance approximation 
\cite{Gerchikov_1997_JPhysB.30.4133, Gerchikov_1998_JPhysB.31.3065, Gerchikov_1997_JPhysB.30.5939}, 
the differential inelastic scattering cross section of fast electrons in collision with fullerenes can 
be defined as a sum of three contributions (we use the atomic system of units, $m_e = |e| = \hbar = 1$):
\begin{equation}
\frac{\d^3\sigma}{\d\E_2 \d\Om_{{\bfp}_2}} = \frac{\d^3\sigma^{(v)} }{\d\E_2 \d\Om_{{\bfp}_2}} +
\frac{\d^3\sigma^{(s_1)} }{\d\E_2 \d\Om_{{\bfp}_2}} + \frac{\d^3\sigma^{(s_2)} }{\d\E_2 \d\Om_{{\bfp}_2}} \ ,
\label{Equation.01}
\end{equation}
where
\begin{eqnarray}
\left\{
\begin{array}{l l}
\displaystyle{ \frac{\d^3\sigma^{(v)} }{\d\E_2 \d\Om_{{\bfp}_2}} } = \frac{2 R_2 p_2}{\pi q^4 p_1} \, \om
\sum\limits_{l} \frac{ \om_p^2\, \Gamma_l^{(v)}\, V_l(q) }
{ \bigl(\om^2-\om_p^2\bigr)^2+\om^2\Gamma_l^{(v)2} }
\vspace{0.2cm} \\
\displaystyle{ \frac{\d^3\sigma^{(s_1)} }{\d\E_2 \d\Om_{{\bfp}_2}} } = \frac{2 R_2 p_2}{\pi q^4 p_1} \, \om
\sum\limits_{l} \frac{ \om_{1l}^2\, \Gamma_{1l}^{(s)}\, S_{1l}(q) }
{ \bigl(\om^2-\om_{1l}^2\bigr)^2+\om^2\Gamma_{1l}^{(s)2}}
\vspace{0.2cm} \\
\displaystyle{ \frac{\d^3\sigma^{(s_2)} }{\d\E_2 \d\Om_{{\bfp}_2}} } = \frac{2 R_2 p_2}{\pi q^4 p_1} \, \om
\sum\limits_{l} \frac{ \om_{2l}^2\, \Gamma_{2l}^{(s)}\, S_{2l}(q) }
{ \bigl(\om^2-\om_{2l}^2\bigr)^2+\om^2\Gamma_{2l}^{(s)2} }
\end{array} \right.
\label{Equation.02}
\end{eqnarray}
are obtained within the plane-wave first Born approximation.
Here $\E_2 = {\bfp}_2^2/2 $ is the kinetic energy of the scattered electron, 
$\Om_{{\bfp}_2}$ its solid angle, 
${\bfp}_1$ and ${\bfp}_2$ the initial and the final momenta of the projectile electron, 
${\bfq} = {\bfp}_1 - {\bfp}_2$ the transferred momentum, 
$\om = \E_1 - \E_2$ the energy loss and $\E_1$ the kinetic energy of the incident electron. 
$\om_p = \sqrt{3N/(R_2^3 - R_1^3)}$ is the volume plasmon frequency ($N$ stands for a number of 
delocalized electrons in the fullerene), 
$\om_{1l}$ and $\om_{2l}$ are the frequencies of the symmetric and antisymmetric surface plasmons 
of multipolarity $l$ \cite{Oestling_1993_EurophysLett.21.539}:
\begin{equation}
\frac{\om_{(1,2)l}^2}{\om_p^2} = 
\frac12 \mp  \frac{1}{2(2l+1)} \sqrt{1 + 4l(l+1) \left(R_1 / R_2 \right)^{2l+1}} \ ,
\label{MultipoleVariation.2a}
\end{equation}
where the signs ''$-$'' and ''$+$'' correspond to the symmetric ($\om_{1l}$) and the antisymmetric 
($\om_{2l}$) modes, respectively. 
$\Gamma_l^{(v)}$ and $\Gamma_{jl}^{(s)}\, (j = 1,2)$ are the widths of the plasmon excitations. 
Functions $V_l(q)$, $S_{1l}(q)$ and $S_{2l}(q)$ are the diffraction factors depending on the 
transferred momentum $q$. 
They determine the relative contribution of the multipole plasmon modes in various ranges of electron 
scattering angles and, thus, the resulting shape of the differential energy loss spectrum. 
Explicit expressions for these functions are presented in \cite{Plasmons_formalism_2012}.

The introduced model is applicable within the long wavelength limit, when the characteristic 
scattering length, $1/q$, is large. 
Under the condition of the small transferred momentum $q$, the volume plasmon is characterized by the 
constant frequency $\om_p$ which does not depend on the transferred momentum \cite{Landau_Lifshitz_10}.
In this paper, we do not consider the dependence of the plasmon widths on the transferred momentum 
which was studied in \cite{Gerchikov_2000_PhysRevA.62.043201}. 
The widths are treated as external parameters which are not calculated within the present model.

Assuming $R_1 \to R_2 \equiv R$, we come to the model which treats a fullerene as an infinitely 
thin sphere. 
In this limit, the cross section, $\d^3\sigma \equiv \d^3\sigma^{(s_1)}$, is defined only by the 
single surface plasmon, and (\ref{Equation.01}) and (\ref{Equation.02}) transform into the 
following expression \cite{Gerchikov_1997_JPhysB.30.4133,Gerchikov_1998_JPhysB.31.3065,
Mikoushkin_1998_PhysRevLett.81.2707}:
\begin{equation}
\frac{\d^3\sigma}{\d\E_2 \d\Om_{{\bfp}_2}} =
\frac{4 R\, p_2}{\pi q^4 p_1} \, \om
\sum_{l}
{ (2l+1)^2\, \om_{l}^2\, \Gamma_{l}\, j_l^2(qR)
\over\bigl(\om^2-\om_{l}^2\bigr)^2 + \om^2\Gamma_{l}^{2}}\,
 \ ,
\label{Equation.03}
\end{equation}
where $\om_{l} = \sqrt{l(l+1) N /(2l+1) R^3}$ is the surface plasmon frequency and 
$\Gamma_{l}\equiv \Gamma_{1l}^{(s)}$ is its width.

According to \cite{Gerchikov_1997_JPhysB.30.4133}, we can estimate the angular momentum range 
which should be considered within the model. 
In the case of the C$_{60}$ fullerene, multipole excitations with $l > 3$ are formed by single-electron 
transitions rather than collective electron excitations, so only terms corresponding to the dipole 
$(l = 1)$, quadrupole $(l=2)$ and octupole $(l=3)$ plasmon excitations should be included to the 
sum over $l$ in (\ref{Equation.02}) and (\ref{Equation.03}). 
The diffraction phenomena observed in \cite{Mikoushkin_1998_PhysRevLett.81.2707} manifest themselves 
in the dominating contribution of different multipolar excitations at different electron scattering 
angles \cite{Plasmons_experiment_2012}.

The theory presented in the paper is based on the model assumptions governed by two parameters: 
the number of multipole terms to be accounted for ($l_{\rm max} = 3$) and the widths of plasmon 
resonances, $\Gamma_{jl}^{(s)}$ and $\Gamma_l^{(v)}$. 
The choice of these parameters is limited by their physical meaning, and thus, allows one to vary 
the overall behavior of the inelastic scattering cross section only a little. 
These parameters can be chosen on the basis of data available in literature according to their actual 
physical limitations. 
The concrete calculation of the values of these parameters can be done, but it is beyond the topic of 
this paper, and will be a subject for further investigations.

The main advantage of the method developed is that it provides a clear explanation of the main 
resonance-like structure of the inelastic scattering cross section on the basis of the major physical 
phenomena at play, namely by the excitation of multipole plasmon resonances by electron impact.
More accurate calculations based on the explicit use of quantum-mechanical methods should reveal a 
further, more detailed structure of the cross section. 
Therefore, the plasmon resonance approximation utilized and further advanced in the present work turns 
out to be also a very useful tool for the interpretation of the results of such quantum simulations. 
The validity of the plasmon resonance approximation was proved by the comparison of its results with 
those following from {\it ab initio} quantum calculations for the case of metal clusters \cite{Gerchikov_2000_PhysRevA.62.043201, Gerchikov_1997_JPhysB.30.5939}, and also by the comparison with 
available experimental data \cite{Gerchikov_1997_JPhysB.30.4133, Mikoushkin_1998_PhysRevLett.81.2707}.


A crossed-beam apparatus \cite{Avaldi_1993_PhysRevA.48.1195} has been used to measure the energy 
loss spectrum of C$_{60}$. 
The vacuum chamber contains an electron gun, two twin 180$^{\circ}$ hemispherical electrostatic 
analyzers, rotatable independently in the scattering plane and a resistively heated, anti-inductively 
wounded oven to produce the C$_{60}$ beam \cite{Bolognesi_2008_JPhysB.41.015201}. 
The typical operational temperature of the oven was about 500$^{\circ}$C. 
The compensation of the earth magnetic field has been performed by three pairs of orthogonal square 
coils \cite{Avaldi_1993_PhysRevA.48.1195} external to the vacuum chamber and by an internal 0.4 mm 
thick Skudotech layer \cite{Bolognesi_2008_JPhysB.41.015201}.

The scattered electrons have been analyzed in energy by one of the two electron spectrometers. 
A three-element electrostatic lens focuses the electrons from the target region onto the entrance 
slit of the hemispherical analyser (60 mm mean radius). 
After the angle and energy selection, the electrons have been detected by a channeltron electron multiplier. 
In these experiments the 1000 eV scattered electrons have been slowed down to a pass energy of 50 eV. 
The energy resolution was 1.2 eV full width half maximum, as measured by detecting the elastic scattered 
electrons, and the angular resolution about $\pm 2^{\circ}$. 
The output signals of the detector have been sent to a PCI-6024E National Instruments card through a 
preamplifier and a constant fraction discriminator. 
The typical incident current, monitored by a Faraday Cup, has been in the range of a few $\mu$A. 
A personal computer via a Labview software scans the energy of the incident beam, changing the energy loss 
of the scattered electron, controls the movement of the turntables, sets the dwell time of the measurements, 
monitors the current of the beam during the acquisition and stores the results. 
The scattered angle scale has been calibrated by checking the symmetry of the measured yield at a fixed 
energy loss with respect to the direction of the primary beam, while the zero of the energy loss scale 
has been defined by elastically scattered electrons. 
In the present measurements, the energy of the scattered electrons and the scattering angle have been fixed, 
while the incident energy has been varied to cover the energy loss region of interest.


\begin{figure}
\centering
\includegraphics[scale=0.35,clip]{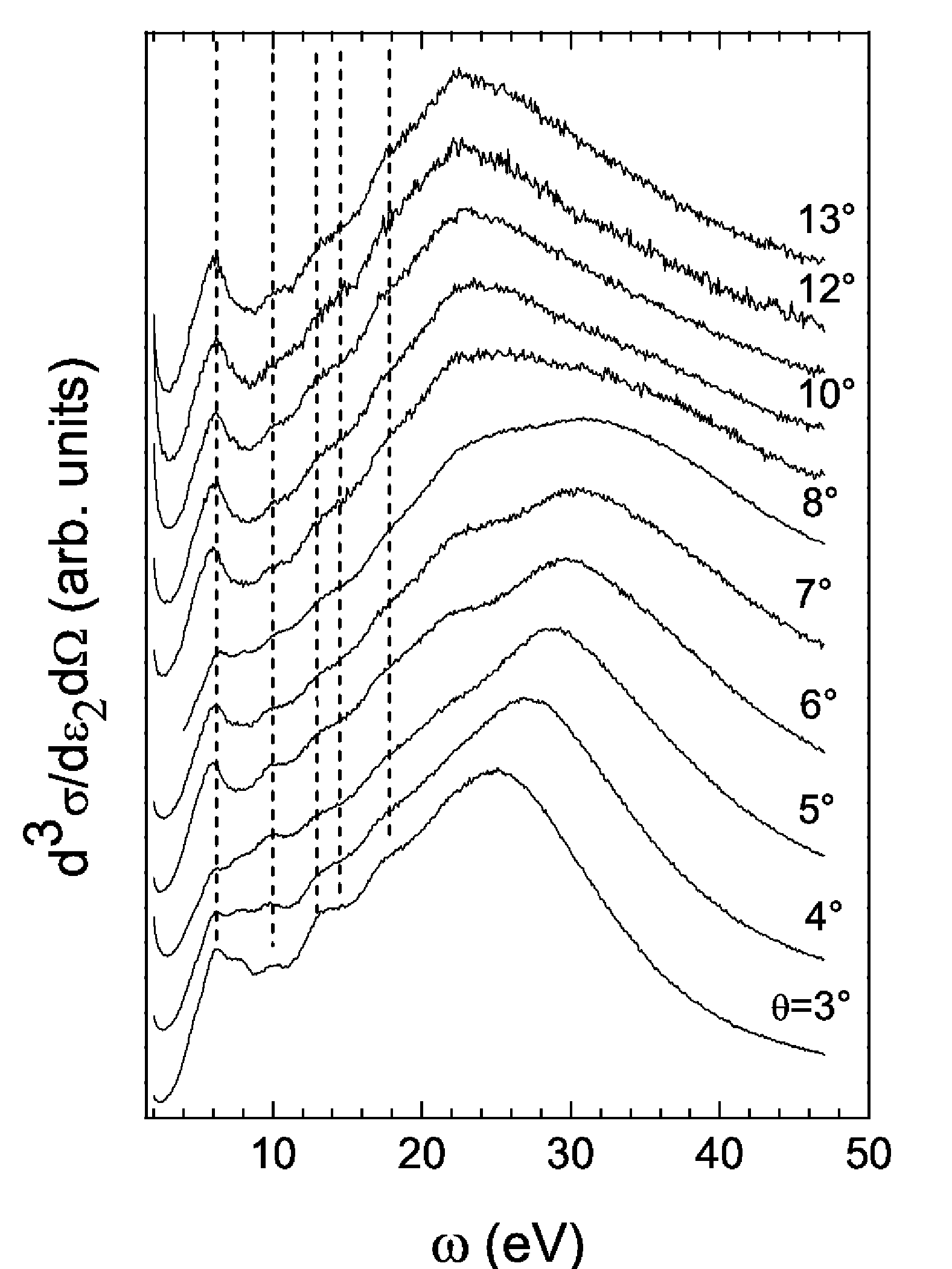}
\caption{
The full set of the electron energy loss spectra measured for the incident energy range 1002-1050 eV and 
for the scattering angle range $\theta = 3^{\circ}\dots13^{\circ}$. 
A broad feature above 20 eV is assigned to the collective excitation of the delocalized $(\sigma+\pi)$ electrons. 
Dashed lines indicate the position of the $\pi$-plasmon and a number of single-electron excitations.}
\label{figure1}
\end{figure}

A series of energy loss spectra, EELS, measured in the scattering angle range 3$^{\circ}\dots13^{\circ}$ 
are presented in figure \ref{figure1}.
For the sake of convenience, all spectra are normalized to 1.
The spectra show a series of structures, starting at about $\om=5$ eV, overimposed to a broad feature 
peaking above 20 eV.
The peaks below 20 eV corresponding to the electronic excitations are marked by dashed lines in figure \ref{figure1}. 
A prominent peak at about 6 eV is consistent with those measured in solids by Lukas \textit{et al}. \cite{Lukas_1992_PhysRevB.45.13694} and observed in the gas phase C$_{60}$ by Keller and Coplan \cite{Keller_Coplan_1992_ChemPhysLett.193.89}. 
According to Barton and Eberlein \cite{Barton_1991_JChemPhys.95.1512} and Bertsch \textit{et al.} \cite{Bertsch_1991_PhysRevLett.67.2690}, this feature is assigned to the plasmon arising from the 
collective motion of $\pi$ electrons.
The sequence of the peaks in the range $10\dots17$ eV has been assigned \cite{Lukas_1992_PhysRevB.45.13694} 
to electronic excited states converging to different ionization bands observed in photoemission \cite{Liebsch_1995_PhysRevA.52.457} and ($e,2e$) experiments \cite{Bolognesi_2011_JPhysConfSer.288.012006}.
Above 20 eV the differential cross section, $\d^3\sigma/\d\E_2 \d\Om_{{\bfp}_2}$, displays a broad 
feature which may be assigned to the $(\sigma + \pi)$-plasmon formed by the collective excitation 
of the delocalized electrons \cite{Oestling_1993_EurophysLett.21.539,Oestling_1996_JPhysB.29.5115}.
With increasing the scattering angle, the shape of the energy loss spectra varies significantly and 
the existence of two different peaks between 20 and 30 eV is clearly seen for the angles 
$\theta = 6^{\circ}\dots8^{\circ}$.

In this paper we focus on the study of the behavior of the main peak above 20 eV.
To understand the reason of the EELS shape variation we calculated the differential cross section 
within the three-plasmon model (see (\ref{Equation.01}) and (\ref{Equation.02})).
Radius of the C$_{60}$ molecule is equal to 3.54 \AA, and the width of the spherical shell is set 
to 1.5 \AA, which was obtained by R\"udel \emph{et al}. \cite{Ruedel_2002_PhysRevLett.89.125503}.
The ratio $\gamma_{1l}^{(s)} = \Gamma_{1l}^{(s)}/\om_{1l}$ for the symmetric mode is equal to 0.6 
which is close to the experimental values obtained from the photoionization and energy loss experiments 
on neutral C$_{60}$ \cite{Hertel_1992_PhysRevLett.68.784, Mikoushkin_1998_PhysRevLett.81.2707}. 
The value $\gamma_{1l}^{(s)} =0.6$ is chosen according to the previous theoretical investigations of the 
electron energy loss based on the plasmon resonance approximation \cite{Gerchikov_1997_JPhysB.30.4133,Gerchikov_1998_JPhysB.31.3065}.
The widths of the antisymmetric mode as well as of the volume plasmon were varied to obtain a better 
agreement with the experimental data. 
In the present calculations, the ratios 
$\gamma_{2l}^{(s)} = \Gamma_{2l}^{(s)}/\om_{2l}$ and $\gamma_{l}^{(v)} = \Gamma_{l}^{(v)}/\om_{p}$ are 
equal to 1.
The value $\gamma_{2l}^{(s)} =1.0$ corresponds to the widths of the second plasmon resonance obtained by 
Scully et al. \cite{Scully_2005_PhysRevLett.94.065503} studying photoionization of C$_{60}^{q+}$ ($q = 1-3$) ions.
In the case of the infinitely thin fullerene, we used the ratio $\gamma_{l} \equiv \gamma_{1l}^{(s)} = 0.6$ 
according to \cite{Gerchikov_1997_JPhysB.30.4133,Gerchikov_1998_JPhysB.31.3065}.

\begin{figure}
\centering
\includegraphics[scale=0.5,clip]{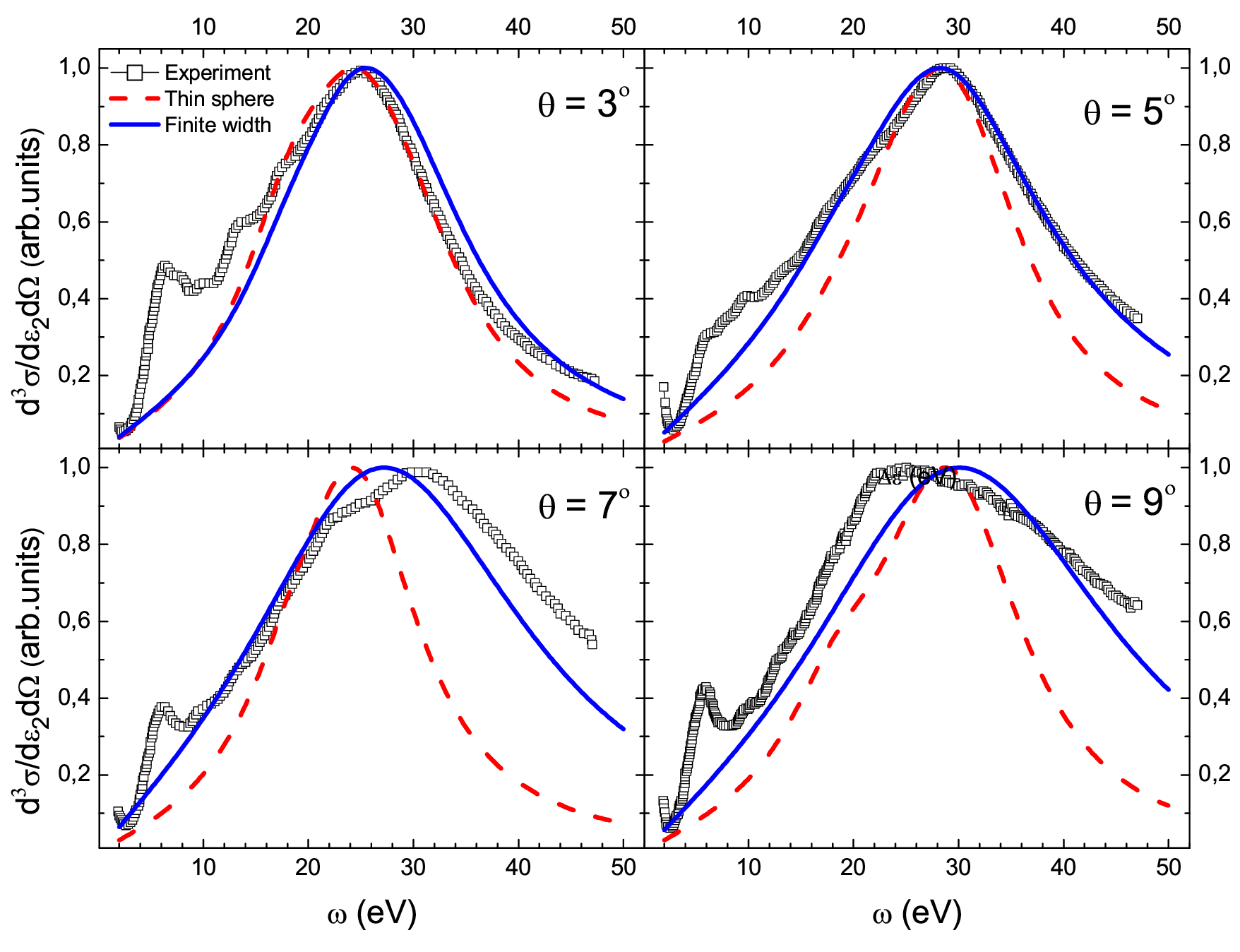}
\caption{
Comparison of the experimental EELS with theoretical results obtained within the three-plasmon model 
and the single-plasmon model for the scattering angles $\theta = 3^{\circ} \dots 9^{\circ}$. 
Black squares represent the experimental data, cross section obtained for the case of the infinitely 
thin fullerene is presented by the dashed (red) line, the solid (blue) line denotes the cross section 
for a fullerene with the finite width.}
\label{figure2}
\end{figure}

The comparison of the experimental data with the model calculations for the scattering angles 
$\theta = 3^{\circ} \dots 9^{\circ}$ is presented in figure \ref{figure2}. 
Both the experimental and the theoretical curves are normalized to 1.
Black squares represent the experimental data, differential cross section obtained for the case 
of the infinitely thin fullerene is presented by the dashed (red) line while the solid (blue) line 
denotes the cross section for a fullerene with the finite width.
At the small scattering angle, $\theta = 3^{\circ}$, both models show quite good agreement with 
the experimental curve.
It means that in this case the symmetric mode of the surface plasmon dominates the cross section, 
while the influence of the two other plasmons is rather weak.
Increasing the scattering angle, the cross section formed by the single surface plasmon 
(dashed red line) becomes much narrower than the experimental spectrum.
On the other hand, the three-plasmon model (solid blue line) leads to a good quantitative agreement 
with the experiment.
It means that for the scattering angle larger than $\theta = 5^{\circ}$, the second surface plasmon 
and the volume plasmon begin to play a significant role in the formation of the energy loss spectrum. 
They may also contribute to the formation of the two resonances between 20 and 30 eV at the scattering 
angles $\theta = 6 \dots 8^{\circ}$ (see figure \ref{figure1}).
As seen from figure \ref{figure2}, at $\theta = 7^{\circ}$ the width of the plasmon resonance 
calculated within the three-plasmon model (solid blue line) corresponds to the total width of 
the experimental broad structure but does not reproduce the two more narrow individual peaks. 
Refining further the theory presented in this paper, one should be able to reproduce also the details of 
the two-resonant structure appearing between 20 and 30 eV at the angles $\theta = 6 \dots 8^{\circ}$.
This is one of the questions for a further investigation.

\begin{figure}
\centering
\includegraphics[scale=0.5,clip]{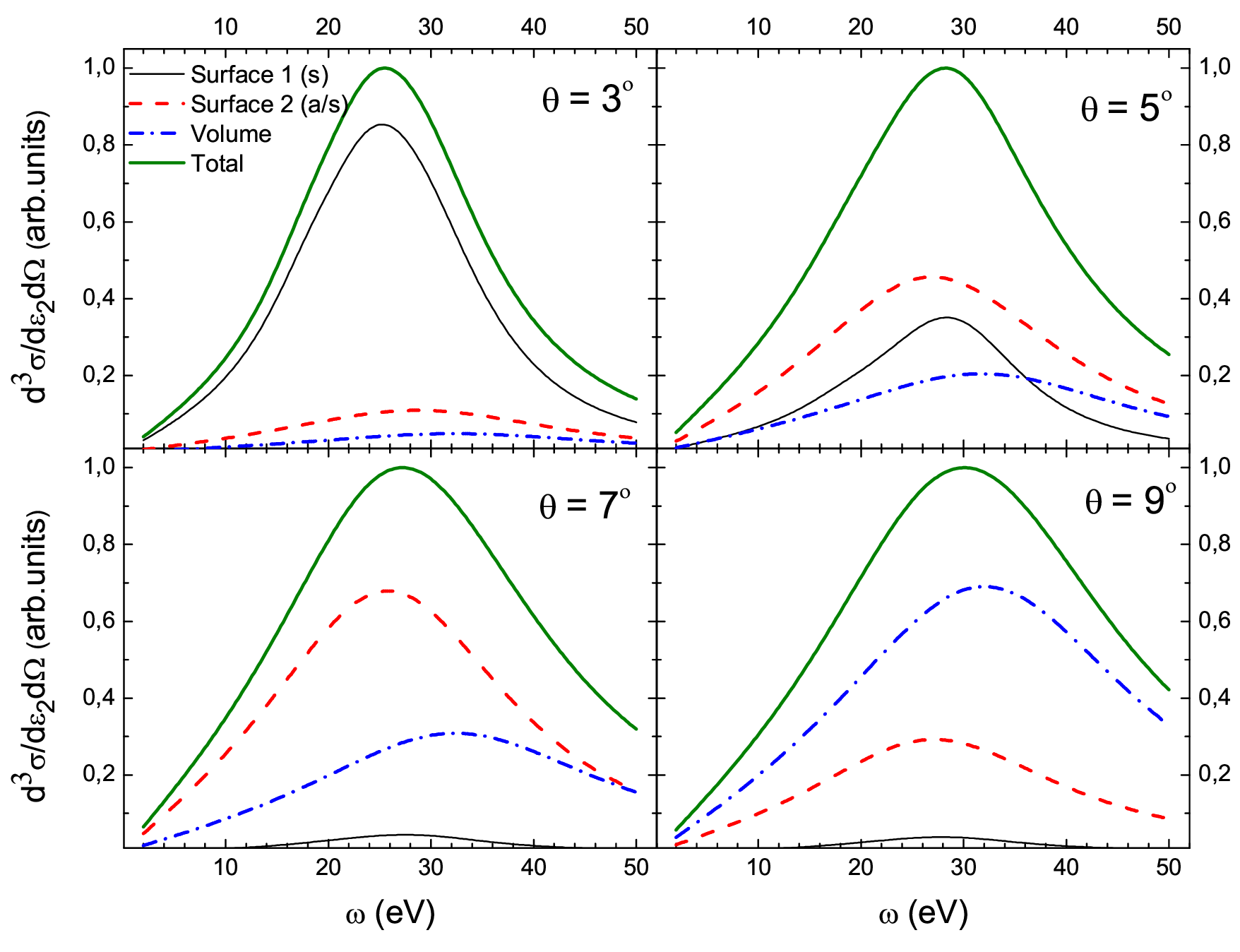}
\caption{
The total cross section and partial contributions of the volume and the two surface plasmons for the 
scattering angles $\theta = 3^{\circ} \dots 9^{\circ}$.
The symmetric (s) and the antisymmetric (a/s) modes of the surface plasmon are shown by the thin 
solid black line and the dashed red line, respectively; the volume plasmon contribution is shown by 
the dash-dotted blue line. The total cross section is shown by the thick green line.}
\label{figure3}
\end{figure}

A few words should be said about possible thermal effects on the energy loss spectra. 
{\it Ab initio} calculations of atomic clusters based on the dynamical jellium model 
\cite{Gerchikov_2000_JPhysB.33.4905} revealed the broadening of the linewidths of single-electron 
excitations in the vicinity of the plasmon resonance due to the electron coupling with the surface 
and volume vibrational modes of the ionic background. 
For the case of the Na$_{40}$ cluster it was shown that the broadening width increases slightly with 
the temperature and is about 0.15 eV for $T = 400$ K \cite{Gerchikov_2000_JPhysB.33.4905}. 
For the C$_{60}$ fullerene, the additional broadening of the spectra should be of the similar order of 
magnitude, therefore possible thermal effects due to the operational temperature of 500$^{\circ}$C 
should not influence significantly the energy loss spectra. 
The multifragmentation process does not play a role at this temperature, since the multifragmentation 
of C$_{60}$ in C$_2$ dimers or other small carbon fragments occurs at significantly higher temperatures, 
being above 3000~K \cite{Hussien_2010_EPJD.57.207}.

To illustrate a relative role of each of the three plasmons in the formation of the energy loss 
spectrum, we present in figure \ref{figure3} the calculated partial contribution of each plasmon 
(\ref{Equation.02}) as well as their sum for the scattering angles in the range 
$\theta = 3^{\circ}\dots9^{\circ}$.
At the small scattering angle, $\theta = 3^{\circ}$, the symmetric mode of the surface plasmon 
(thin solid black line) dominates the cross section.
A similar behavior was revealed by Scully \textit{et al}. \cite{Scully_2005_PhysRevLett.94.065503} 
in the photoionization. 
In fact, in the case of the uniform external field ($q \to 0$), there is no volume plasmon excitation 
in the system and the symmetric plasmon mode exceeds significantly the antisymmetric mode.
Non-uniformity of the external field causes the formation of the volume plasmon whose contribution to 
the cross section is insignificant when the scattering angle is small.
With increasing the scattering angle ($\theta = 5^{\circ}$ and $7^{\circ}$), the symmetric mode of the 
surface plasmon becomes less relevant and the antisymmetric mode (dashed red line) more prominent.
At the larger angle ($\theta = 9^{\circ}$), the symmetric surface plasmon almost does not contribute 
to the cross section while the volume plasmon (dash-dotted blue line) becomes dominant. 
Thereby, the origin of the two peaks in the energy loss range from 20 to 30 eV at the angles 
$\theta = 6^{\circ}$\dots $8^{\circ}$ can be explained by the contribution of the antisymmetric surface 
and the volume plasmons.

The peak position of each plasmon resonance (\ref{Equation.02}) as well as of the resulting cross 
section $\d^3\sigma/\d\E_2 \d\Om_{{\bfp}_2}$ is influenced by the manifestation of the diffraction effects.
It was shown \cite{Mikoushkin_1998_PhysRevLett.81.2707} that plasmon modes with different angular 
momenta provide dominating contributions to the differential cross section at different scattering angles, 
which leads to the significant angular dependence of the energy loss spectrum. 
This phenomenon was described in terms of the electron diffraction at the fullerene edge \cite{Mikoushkin_1998_PhysRevLett.81.2707}.
As seen from (\ref{Equation.02}), the resonance peak of each plasmon is defined not only by the plasmon 
frequencies $\om_p$, $\om_{1l}$ and $\om_{2l}$ but also by the multipolar diffraction factors 
$V_l(q)$, $S_{1l}(q)$ and $S_{2l}(q)$ which depend on the transferred momentum $q$. 
In the limiting case of an infinitely thin layer, this dependence is described by the spherical Bessel 
functions $j_l^2(qR)$ (see (\ref{Equation.03})) which oscillate with $q$ and, thus, give suppression 
and enhancement of the partial plasmon modes at certain angles \cite{Mikoushkin_1998_PhysRevLett.81.2707}. 
The incident energy of the projectile does not influence on this behavior and defines only the absolute 
value of the cross section.


To conclude, we performed a joint experimental and theoretical investigation of collective excitations 
in C$_{60}$ in the process of inelastic scattering of electrons in collision with fullerenes.
An extensive set of measurements of the energy loss spectrum of the C$_{60}$ molecule has been performed 
for the scattering angle range from $3^{\circ}$ to $13^{\circ}$.
We have introduced a new theoretical model which accounts for the two modes of the surface plasmon 
as well as the volume plasmon, when the fullerene is modeled as a spherical shell of a finite width.
Theoretical results obtained within this model are in a good agreement with the experimental data.
The present results show that collective excitations provide the main contribution to the inelastic 
scattering cross section of electrons over a broad energy range and, as opposed to the photoionization, 
both the surface plasmons as well as the volume one contribute to the cross section. 
It has been shown that the symmetric mode of the surface plasmon dominates at smaller scattering angles, 
while at larger angles the antisymmetric and the volume plasmons make the most prominent contribution.

\ack

The work was partially supported by PRIN 2009SLKFEX and 2009W2W4YF. 
A.V.V. is grateful to DAAD for financial support.

\section*{References}


\begin{thebibliography}{99}

\bibitem{Eisenberg_Greiner}
  Eisenberg J M and Greiner W 1970
  {\it Nuclear Models: Collective and Single-Particle Phenomena}
  (Amsterdam: North-Holland)

\bibitem{Connerade_GR}
  Connerade J P, Esteva J M and Karnatak R C 1987
  {\it Giant Resonances in Atoms, Molecules, and Solids}
  (New York: Plenum Press)

\bibitem{deHeer_1993_RevModPhys.65.611}
  de Heer W A 1993 {\it Rev. Mod. Phys.} {\bf 65} 611

\bibitem{Kreibig_Vollmer}
  Kreibig U and Vollmer M 1995
  {\it Optical Properties of Metal Clusters}
  (Berlin: Springer-Verlag)

\bibitem{Bohm_Pines_1952_PhysRev.85.338}
  Pines D and Bohm D 1952 {\it Phys. Rev.} {\bf 85} 338

\bibitem{Kaplan_1987_AdvChemPhys.68.225}
  Kaplan I G and Miterev A M 1987 {\it Adv. Chem. Phys.} {\bf 68} 225

\bibitem{Solovyov_review_2005_IntJModPhys.19.4143}
  Solov'yov A V 2005 {\it Int. J. Mod. Phys.} B {\bf 19} 4143

\bibitem{Phaneuf_2010_Plasmon_in_Fullerenes}
  Phaneuf R A 2010 {\it Handbook of Nanophysics: Clusters and Fullerenes}
  ed K D Sattler (Boca Raton: CRC Press) p 35-1

\bibitem{Singhal_2008_ApplPhysLett.93.103114}
  Singhal R {\it et al} 2008 {\it Appl. Phys. Lett.} {\bf 93} 103114

\bibitem{Porter_Gass_2008}
  Porter A and Gass M 2008
  {\it Medicinal Chemistry and Pharmacological Potential of Fullerenes and Carbon Nanotubes}
  vol 1, ed F Cataldo T and Da Ros (New York: Springer Science+Business Media B.V.) p 267

\bibitem{Gerchikov_2000_PhysRevA.62.043201}
  Gerchikov L G, Ipatov A N, Polozkov R G and Solov'yov A V 2000 
  {\it Phys. Rev.} A {\bf 62} 043201

\bibitem{Bertsch_1991_PhysRevLett.67.2690}
  Bertsch G F, Bulgac A, Tomanek D and Wang Y 1991 
  {\it Phys. Rev. Lett.} {\bf 67} 2690


\bibitem{Hertel_1992_PhysRevLett.68.784}
  Hertel I V, Steger H, de Vries J, Weisser B, Menzel C, Kamke B and Kamke W 1992 
  {\it Phys. Rev. Lett.} {\bf 68} 784

\bibitem{Keller_Coplan_1992_ChemPhysLett.193.89}
  Keller J W and Coplan M A 1992 
  {\it Chem. Phys. Lett.} {\bf 193} 89

\bibitem{Reinkoester_2004_JPhysB.37.2135}
  Reink\"oster A, Korica S, Pr\"umper G, Viefhaus J, Godehusen K, Schwarzkopf O, Mast M and Becker U 2004
  {\it J. Phys. B: At. Mol. Opt. Phys.} {\bf 37} 3125

\bibitem{Scully_2005_PhysRevLett.94.065503}
  Scully S W J {\it et al} 2005 
  {\it Phys. Rev. Lett.} {\bf 94} 065503

\bibitem{Korol_AS_2007_PhysRevLett_Comment}
  Korol A V and Solov'yov A V 2007 
  {\it Phys. Rev. Lett.} {\bf 98} 179601

\bibitem{Scully_2007_PhysRevLett_Reply}
  Scully S W J {\it et al} 2007 {\it Phys. Rev. Lett.} {\bf 98} 179602

\bibitem{Gerchikov_1997_JPhysB.30.4133}
  Gerchikov L G, Solov'yov A V, Connerade J-P and Greiner W 1997
  {\it J. Phys. B: At. Mol. Opt. Phys.} {\bf 30} 4133

\bibitem{Gerchikov_1998_JPhysB.31.3065}
  Gerchikov L G, Ipatov A N, Solov'yov A V and Greiner W 1998
  {\it J. Phys. B: At. Mol. Opt. Phys.} {\bf 31} 3065

\bibitem{Mikoushkin_1998_PhysRevLett.81.2707}
  Gerchikov L G, Efimov P V, Mikoushkin V M and Solov'yov A V 1998
  {\it Phys. Rev. Lett.} {\bf 81} 2707

\bibitem{PuskaNiemenen_1993_PhysRevA.47.1181}
  Puska M J and Nieminen R M 1993 {\it Phys. Rev.} A {\bf 47} 1181

\bibitem{Oestling_1993_EurophysLett.21.539}
  \"Ostling D, Apell P and Rosen A 1993 {\it Europhys. Lett.} {\bf 21} 539

\bibitem{Ruedel_2002_PhysRevLett.89.125503}
  R\"udel A, Hentges R, Becker U, Chakraborty H S, Madjet M E and Rost J-M 2002
  {\it Phys. Rev. Lett.} {\bf 89} 125503

\bibitem{Lambin_Lukas_1992_PhysRevB.46.1794}
  Lambin P, Lucas A A and Vigneron J-P 1992 {\it Phys. Rev.} B {\bf 46} 1794

\bibitem{Lo_2007_JPhysB.40.3973}
  Lo S, Korol A V and Solov'yov A V 2007
  {\it J. Phys. B: At. Mol. Opt. Phys.} {\bf 40} 3973

\bibitem{Gerchikov_1997_JPhysB.30.5939}
  Gerchikov L G, Ipatov A N and Solov'yov A V 1997
  {\it J. Phys. B: At. Mol. Opt. Phys.} {\bf 30} 5939

\bibitem{Plasmons_formalism_2012}
  Verkhovtsev A V, Korol A V and Solov'yov A V
  2012 submitted to {\it Eur. Phys. J.} D (see also arXiv:1202.6211v2 [physics.atm-clus])

\bibitem{Landau_Lifshitz_10}
  Lifshitz E M and Pitaevskii L P 1981 {\it Physical Kinetics} ({\it Course of Theoretical Physics} vol 10)
  (New York: Pergamon Press)

\bibitem{Plasmons_experiment_2012}
  Bolognesi P, Avaldi L, Ruocco A, Verkhovtsev A, Korol A V and Solov'yov A V 2012
  submitted to {\it Eur. Phys. J.} D

\bibitem{Avaldi_1993_PhysRevA.48.1195}
  Avaldi L, Camilloni R, Multari R, Stefani G, Zhang X, Walters H R J and Whelan C T 1993
  {\it Phys. Rev.} A {\bf 48} 1195

\bibitem{Bolognesi_2008_JPhysB.41.015201}
  Bolognesi P, Bohachov H, Borovik V, Veronesi S, Flammini R, Fainelli E, Borovik A, 
  Martinez J, Whelan C T, Walters H R J, Kheifets A and Avaldi L 2008
  {\it J. Phys. B: At. Mol. Opt. Phys.} {\bf 41} 015201

\bibitem{Lukas_1992_PhysRevB.45.13694}
  Lucas A, Gensterblum G, Pireaux J J, Thiry P A, Caudano R, Vigneron J P, Lambin P and Kr\"atschmer W 1992
  {\it Phys. Rev.} B {\bf 45} 13694

\bibitem{Barton_1991_JChemPhys.95.1512}
  Barton G and Eberlein C 1991 {\it J. Chem. Phys.} {\bf 95} 1512

\bibitem{Liebsch_1995_PhysRevA.52.457}
  Liebsch T {\it et al.} 1995 {\it Phys. Rev.} A {\bf 52} 457

\bibitem{Bolognesi_2011_JPhysConfSer.288.012006}
  Bolognesi P, Pravi\c{c}a L, Berakdar J, Pavlyukh Y, Camilloni R, Cvejanovic D, Avaldi L 2011
  {\it J. Phys.: Conf. Ser.} {\bf 288} 012006

\bibitem{Oestling_1996_JPhysB.29.5115}
  \"Ostling D, Apell P, Mukhopadhyay G and Rosen A 1996
  {\it J. Phys. B: At. Mol. Opt. Phys.} {\bf 29} 5115

\bibitem{Gerchikov_2000_JPhysB.33.4905}
  Gerchikov L G, Ipatov A N, Solov'yov A V and Greiner W 2000
  {\it J. Phys. B: At. Mol. Opt. Phys.} {\bf 33} 4905

\bibitem{Hussien_2010_EPJD.57.207}
  Hussien A, Yakubovich A V, Solov'yov A V and Greiner W 2010
  {\it Eur. Phys. J.} D {\bf 57} 207

\end{thebibliography}

\end{document}